# Direct evidence of hierarchical assembly at low masses from isolated dwarf galaxy groups

S. Stierwalt[1,2,*], S.E. Liss[2], K.E. Johnson[2], D.R. Patton[3], G.C. Privon[4], G. Besla[5], N. Kallivayalil[2] & M. Putman[6]

[1] National Radio Astronomy Observatory, 520 Edgemont Road, Charlottesville, VA 22904, USA
[2] Department of Astronomy, University of Virginia, 530 McCormick Road, Charlottesville, VA 22904, USA
[3] Department of Physics & Astronomy, Trent University, 1600 West Bank Drive, Peterborough, ON, K9L 0G2, Canada
[4] Instituto de Astrofísica, Facultad de Física, Pontifica Universidad Católica de Chile, Casilla 306, Santiago 22, Chile
[5] Department of Astronomy, University of Arizona, 933 N Cherry Ave, Tucson, AZ 85719, USA
[6] Department of Astronomy, Columbia University, Mail Code 5246, 550 West 120th Street, New York, NY 10027, USA

**The demographics of dwarf galaxy populations have long been in tension with predictions from the Λ Cold Dark Matter (ΛCDM) paradigm[1-4]. If primordial density fluctuations were scale-free as predicted, dwarf galaxies should themselves host dark matter subhaloes[5], the most massive of which may have undergone star formation resulting in dwarf galaxy groups. Ensembles of dwarf galaxies are observed as satellites of more massive galaxies[6-9], and there is observational[10] and theoretical[11] evidence to suggest that these satellites at z=0 were captured by the massive host halo as a group. However, the evolution of dwarf galaxies is highly susceptible to environment[12-14] making these satellite groups imperfect probes of ΛCDM in the low mass regime. We have identified one of the clearest examples to date of hierarchical structure formation at low masses: seven isolated, spectroscopically confirmed groups with only dwarf galaxies as members. Each group hosts 3-5 known members, has a baryonic mass of ~$4.4\times10^9$ to $2\times10^{10}$ M$_\odot$, and requires a mass-to-light ratio of <100 to be gravitationally bound. Such groups are predicted to be rare theoretically and**

**found to be rare observationally at the current epoch and thus provide a unique window into the possible formation mechanism of more massive, isolated galaxies.**

A key prediction of the Lambda Cold Dark Matter paradigm is that the merging of smaller galaxies to form larger ones should occur even at the lowest galaxy masses. Thus, theoretical simulations predict the existence of mergers involving only dwarf galaxies and groups with only dwarf galaxy members. However, less than 5% of dwarf galaxies are observed to have close companions (see Methods), and most galaxy surveys are too shallow to consistently detect such companions even when they are present. Thus, other than a few cases of tantalizing yet circumstantial evidence for dwarf-dwarf mergers[15-17], direct observational evidence of this hierarchical structure formation at the low galaxy mass end has remained mostly elusive. Here we report one of the clearest examples to date: the discovery of isolated groups of galaxies that contain only low mass, dwarf galaxy members and that are consistent with being gravitationally bound. The existence of these dwarf-only groups suggests that, given time, hierarchical merging will turn some of these groups into the isolated, intermediate-mass galaxies more commonly observed at the current epoch.

The seven groups, which each host at least 3 to 5 dwarf galaxy members, were discovered while investigating dwarf-dwarf galaxy interactions in the panchromatic TiNy Titans (TNT) survey[18]. As part of an effort to understand the dwarf-dwarf merger sequence, we searched the spectroscopic catalog of the Sloan Digital Sky Survey (SDSS) for nearby dwarf galaxy pairs in isolation (> 1.5 Mpc from more massive neighbors of $M_* > 5 \times 10^9$ $M_\odot$). The distance of 1.5 Mpc from a massive host marks an important boundary between satellite versus field dwarf galaxy evolution. Beyond 1.5 Mpc, the observed quenched fraction of dwarfs falls to zero[12] indicating the end of the sphere of influence of the massive host. Also beyond 1.5 Mpc, the escape velocity of a galaxy with $M_* \sim 10^{10}$ $M_\odot$ falls below the typical sound speed of 10 km/s in the interstellar medium thus decreasing the possibility for disruption. Our systematic search revealed 60 isolated dwarf pairs caught in a variety of interaction stages.

The SDSS images for all 60 isolated TNT pairs were visually inspected for any potential low mass neighbors with or without SDSS spectroscopic data. Seven candidate groups were identified and then, when SDSS spectra were not available, targeted for follow up optical spectroscopy using the longslit optical spectrograph on the 3.5-meter telescope at the Apache Point Observatory. All candidate group members, which range in stellar mass from $2 \times 10^6$ $M_\odot$ < $M_*$ < $2 \times 10^9$ $M_\odot$, were confirmed to have velocities within 200 km/s of the mean group velocity (see Table 1). We further obtained very narrowband (10 Å) Hα imaging for three of the groups with the Maryland Magellan Tunable Filter Fabry Perot which confirms all suspected members are within ±200 km/s of the original pair. One additional group was observed with the narrowband (50 Å) Hα filter on the Gemini Multi-Object Spectrograph at Gemini-North (see Figure 1).

Although their 3D structure is not known, all seven groups are consistent with being bound structures. First, all observed group members are within roughly a virial radius of the most massive group member (~100 kpc for a halo mass of $10^{11}$ $M_\odot$). Second, by equating the 3D velocity dispersion for each group with its escape velocity and by adopting the largest 2D distance between any two group members as the group size, we determine a lower limit on the total (baryonic + dark matter) mass required for the group to be gravitationally bound. As expected, the total masses required to be gravitationally bound exceed the known baryonic masses for each group. The resulting minimum required total mass-to-light (M/L) ratios range from ~0.7 to 11 $M_\odot/L_\odot$ (see Table 2), which are consistent with the M/L ratios of 10-100 that have previously been determined for individual dwarf galaxies[19,20]. We further use our knowledge of the velocities and projected distances of individual group members (see Figure 2) to determine a projected total mass (baryonic + dark matter) estimate for each group (see Methods). The resulting M/L ratios from those mass predictions range from 12 to 70 $M_\odot/L_\odot$ and thus also suggest the TNT dwarf groups do not require an unusual amount of dark matter to be bound systems.

Without independent distance information, projection effects can affect both our distance and velocity estimates, either of which could lead to either interlopers or unbound groups. For example, large transverse motions could lead to 3D velocities that are too high for the systems to be bound. However, this would only be true for the TNT groups if their 3D velocity dispersions are >10 times those measured in 1D. Further increasing the likelihood that the TNT groups are bound structures, the TNT groups have relative line-of-sight velocity differences between group members consistent with those measured for bound groups found in mock catalogs based on the Millennium-II simulation ($\Delta v_{l.o.s.}$ up to 200-300 km/s for groups with $9 < \log(M_*/M_\odot) < 9.5$)[28]. The observed groups could also be transient systems caught at a moment of passing but not gravitationally bound, distinct systems. An estimated 59% of groups of more massive galaxies identified via their projected radial and velocity separations in the Millennium Simulation are found to be actually physically associated[21]. If this result holds to lower masses, a similar fraction (at least 4 of the 7) of the identified TNT groups are expected to be bound groups. To further probe the three dimensional structure, we have an ongoing project to obtain a low resolution neutral hydrogen gas map for each group to look for bridges, streams, or other features connecting the group members physically to one another.

The compact dwarf groups identified here are distinct from previously known groups and loose associations of dwarf galaxies[22-25] in two primary ways: previously known systems are significantly more extended and/or they are within close proximity to a massive galaxy (< 1.5 Mpc) and thus serve as imperfect laboratories for studying hierarchical structure formation at low masses. Seven associations of dwarf galaxies were identified in the Local Volume (D < 8 Mpc) and are close enough for primary distance measurements using tip of the red giant branch stars[24,25]. These associations all have much larger projected *and* three dimensional sizes compared to the TNT groups (see Figure 2). A consequence of the larger observed sizes are the extremely large M/L ratios (all > 100 with some as high as 700) required for the known dwarf associations to be gravitationally bound. Additionally, two of the seven associations are within 0.6 Mpc of a massive ($M_* > 10^8$ $M_\odot$) host and a third is ~1.2 Mpc from the Milky Way. Another previously known group that more closely resembles the TNT groups is Hickson Compact Group 31 (HCG31). HCG31 includes 7-8 members within ~75 kpc all with stellar masses 1 - 8 × $10^9$ $M_\odot$[26,27]. The five galaxies toward the core of the group are clearly interacting and the group has been identified as a pre-merger. However, at least three massive galaxies reside within 1.5 Mpc of HCG31, the most massive ($\log(M_*/M_\odot)$ = 10.7) at a projected distance of 0.97 Mpc and a velocity separation of only 55 km/s. Thus, the competing larger scale environmental effects (including possible ram-pressure or tidal stripping) may be a challenge to disentangle. The TNT-selected dwarf groups are thus the first detection of dwarf-only groups whose isolation enables us to study hierarchical structure formation at low masses.

ΛCDM not only predicts the existence of dwarf groups but also the statistics of how often they should be observed. According to Millennium-II, dwarf galaxies with $M_* \sim 10^9$ $M_\odot$, the average for the TNT primary (most massive) galaxies, have a 1-3% chance[28] of having a companion of similar mass (i.e. a satellite with a mass ratio of $M_*^{satellite}/M_*^{primary} > 0.2$). This prediction is consistent with the upper limit we estimate for the SDSS spectroscopic catalog on the fraction of dwarfs observed to have a close companion of < 5% (see Methods). The FIRE simulations, which push to even lower primary galaxy masses and focus specifically on isolated environments, produce consistent results[5], specifically that ~5% of isolated dwarf galaxies are found to host a satellite with $M_*^{satellite}/M_*^{primary} \gtrsim 1/3$.

The Millennium-II results further predict that the same TNT primary group members have a 1-10% chance of hosting the second and third most massive satellites observed in the TNT groups (i.e. a satellite with mass ratios of $0.02 < M_*^{satellite}/M_*^{primary} < 0.4$). Of the 60 systematically-selected, isolated TNT pairs, seven have additional observable companions, or 11%. The observed TNT fraction of 11% is consistent with the higher end of the range from predictions from simulations (10%). If only 59% of the TNT groups prove to be bound structures, as observed for more massive galaxies[21], the observed fraction of 11% reduces to ~6% and thus lies right amid the range of predictions.

Thus, multiple theoretical approaches predict that dwarf groups like those presented here, with multiple detectable dwarf members with $M_* > 10^{8-9}$ $M_\odot$, are rare at the current epoch. Below a primary galaxy stellar mass of $10^{10}$ $M_\odot$, the number of satellites is expected to decouple from the stellar mass of the primary galaxy according to the semi-analytic mock catalogs derived from the cosmological Millennium-II simulation[28]. Thus, for primary galaxy masses less than that of the Milky Way (and down to the Millennium-II resolution limit of $M_* \sim 10^6$ $M_\odot$), the number of observed satellites is uniformly low as confirmed by the FIRE simulation results.

Unfortunately, matching observations to theoretical predictions at lower satellite masses than those probed by the more massive TNT group members is currently beyond the reach of large surveys like SDSS. Simulations predict that the fraction of satellites will increase. Again according to Millennium-II, for a primary galaxy mass of $10^9$ $M_\odot$, there is a 40% chance that galaxy will host a satellite with $M_* \sim 10^7$ $M_\odot$ (the lowest mass satellite found in our sample). Results from the FIRE simulations are again consistent with 35% of such primaries hosting an ultrafaint satellite[5] with $M_*^{satellite}/M_*^{primary} \sim 0.005$. Although the TNT groups probe masses as low as $\sim 10^7$ $M_\odot$, the survey is far from complete in this mass regime. The current SDSS spectroscopic catalogs are only complete[29] down to $M_* \sim 10^{9.4}$ $M_\odot$ for galaxies bluer than the green valley (with an even higher mass limit for redder galaxies) and only out to a redshift of z = 0.04.

Thus, a direct comparison to theoretical predictions of the frequency and size of dwarf groups that reliably probe masses near the lower limit and below those of the TNT groups is not yet possible. However, observations have uncovered at least one individual example, that of the dwarf DDO68 and its two stellar streams which are consistent with being cannibalized satellites of even lower mass[15]. Placing this intriguing example into context with more statistical investigations will be fertile ground for future 30-meter class telescopes and deeper, more sensitive surveys like those planned by the Large Synoptic Survey Telescope.

The existence of dwarf-only groups suggests that, given time, hierarchical merging will turn some of these groups into isolated intermediate-mass galaxies. Based on the total stellar masses of the TNT groups (see Table 2), and allowing for some additional conversion of gas into stars, these groups could produce isolated galaxies with masses of ~ $9 < \log(M_*/M_\odot) < 10$. Additionally, a rescaling to lower masses of group merger simulations designed to represent massive galaxies[30] implies that the TNT dwarf groups (if bound) would coalesce into a single remnant within ~1 Gyr. Thus, at least some of the galaxies observed to exist at the current epoch in this mass range may have been built up from mergers within dwarf groups. The TNT dwarf groups presented here provide direct probes of hierarchical structure formation in action at the low mass end, giving us a new window into a process expected to be common at earlier times, but nearly impossible to observe at such redshifts.

## Methods

**Data Reduction** *Optical Imaging with the Maryland Magellan Tunable Filter:* Narrowband H$\alpha$ and continuum observations for three of our candidate groups (dm1049+09, dm1349-02, and dm1623+15) were obtained on April 20-21, 2015 using the Inamori-Magellan Areal Camera & Spectrograph (IMACS) with the Maryland Magellan Tunable Filter (MMTF) on the 6.5m Magellan Baade Telescope. The groups selected for imaging were observable from Magellan in April and within the redshift range of 0.0220 to 0.0549 so that the H$\alpha$ emission line would fall within a single MMTF order-blocking filter centered at 6815 Å with a total available width of ~200 Å. Each field was observed using staring mode with the 10 Å tunable portion of the filter centered on the average wavelength of the H$\alpha$ emission line from the original TNT pair. Images of the continuum in all fields were taken by shifting the filter off of the H$\alpha$ emission by 25-50 Å while avoiding night sky lines. The total exposure time ranged from 80 to 120 minutes for both H$\alpha$ and off-band continuum, split in to individual exposures of 20 minutes each. The individual exposures were dithered by between 30″ and 300″ in order to fill in inter-chip gaps, and care was taken to ensure all targets fell within the Fabry Perot's monochromatic spot. The data were reduced using IRAF and the MMTF pipeline[31], which includes bias subtraction, flat fielding, bad pixel masking (including cosmic rays), sky background subtraction, image registration and chip mosaicing, PSF-matching, and image stacking. Images were flux calibrated using standard stars[32] observed with the same filters as the science targets.

*Optical Imaging with the Gemini Multi-Object Spectrograph:* Hα and r-band continuum observations of an additional group candidate (dm1718+30) were obtained as part of a larger TNT follow up program using the Gemini Multi-Object Spectrograph on the 8.1m Gemini-North Telescope[33]. The field was observed for a total of 460 (280) seconds for Hα (r-band), split in to individual exposures of 115 (70) seconds. Individual exposures were dithered by 5″ in the X- and Y- directions to allow for continuous coverage between chip gaps. The observations were reduced using the Gemini/GMOS IRAF package which includes bias subtracting, flat fielding, chip mosaicing, and image alignment and combining. Flux calibration was done using observations of Landolt standard fields.

*Optical Spectroscopy with the Apache Point Dual Imaging Spectrograph:* Long-slit optical spectra of the group candidate members without SDSS spectroscopic observations were obtained throughout 2015 and 2016 using the Dual Imaging Spectrograph (DIS) on the Apache Point Observatory (APO) 3.5m Telescope. Each target was observed using both the red and blue channels with high resolution R1200 and B1200 gratings and a 2.0″ × 6′ slit. Targets were observed at the parallactic angle with total exposure times ranging from 25 to 160 minutes and individual exposure times ranging between 5 and 10 minutes depending on weather conditions. The red (blue) channel was centered at the approximate wavelength of Hα (Hβ) of the group members that were already confirmed via SDSS spectroscopy. The spectra were reduced using standard IRAF tools, including bias subtraction, scattered light correction, and flat fielding. Wavelength calibration was done using Helium, Neon, and Argon arc lamps. Velocities are measured from Gaussian fits to the Hα emission line and the uncertainties derived from the fit parameters are very low (~1-2 km/s) due to the high resolution spectra.

**Mass, mass-to-light ratio, size, and velocity dispersion calculations** Stellar masses were derived based on SDSS ugriz photometry as described for the original TNT pairs[18]. The brightness of each dwarf is measured by summing the flux within a contour tracing the 2-σ noise level (usually 0.03 to 0.04 nanomaggies) as determined by the r-band image. The same aperture is applied to the remaining four bands and the Galactic dust maps from the COBE/DIRBE satellite[34] are used to determine the galactic extinction correction to be applied along each line of sight and for each SDSS filter. Stellar masses were then derived by taking advantage of the SED fitting built into the kcorrect software[35] but without applying a k-correction to our (low redshift) dwarfs. The software assumes a Chabrier initial mass function[36]. Group neutral masses (fifth column of Table 2) were derived from the single dish Green Bank Observatory 21-cm line spectrum[18]. At L-band, the Green Bank beam size is large (~9′) and thus covers the entire group in each case.

We use the projected mass estimator method[37] to estimate the total mass (baryonic and dark matter) present in each group. More specifically:

$$M = \frac{f_{pm}}{G\,(N - \alpha)} \sum_i^N R_{p,i} \Delta V_i^2$$

where $R_p$ is the 2-D projected distance between an individual group member and the (unweighted) center of the group, $\Delta V$ is the difference in the line-of-sight velocity between an individual group member and the mean group velocity, and the sum for each group is performed over all group members given in Table 1. The constant values used are the total number of group members $N$, $f_{pm} = 20/\pi$, the gravitational constant $G$, and $\alpha = 1.5$ as adopted in the literature[23,37].

The 2-D projected size for each group given in the seventh column of Table 2 represents the largest 2-D projected distance between any two group members. Sizes are calculated by placing all group members at the adopted distance for the group (second column of Table 2) which is the average distance to all members based on line of sight optical velocities and an assumption of Hubble flow. We adopt the value $H_0 = 69$ km s$^{-1}$ Mpc$^{-1}$.

The 2-D projected sizes for the TNT dwarf groups and for groups in the literature are also shown in the right panel of Figure 2. We note that additional group members have been proposed for some of the literature groups - most notably the Antlia B dwarf in the NGC3109 group[38,39] but for consistency we include only originally identified the group members[25].

The 3-D velocity dispersion $\sigma_{3D}$ for each group (sixth column in Table 2) is calculated using the line-of-sight velocities of all group members identified in Table 1 and corrected for potential transverse motions. More specifically:

$$\sigma_{3D} = \sqrt{3} \times \sqrt{\langle v^2 \rangle - \langle v \rangle^2}$$

where $v$ is the array of line-of-sight velocities for all group members.

Both mass-to-light ratios in Table 2 use the total B-band luminosity determined by converting fluxes from SDSS g- and r- band photometry to B-band fluxes[40]:

$$m_B = m_g + 0.2354 + 0.3915(m_g - m_r - 0.6102)$$

where $m_B$, $m_g$, and $m_r$ are the summed apparent magnitudes of the dwarf galaxy group members in B-, g-, and r-band respectively.

The value $(M/L_B)_{bound}$ (eighth column of Table 2) represents the lower limit on the mass-to-light ratio required for the group to be a gravitationally bound structure. Thus, the mass used to calculate $(M/L_B)_{bound}$ is determined by setting the escape velocity of the group equal to the 3-D velocity dispersion. In each case, the minimum mass required for the group to be gravitationally bound exceeds the known mass of baryonic matter (stellar and HI gas). The total mass-to-light ratios suggested by these mass estimates are further consistent with those typically observed for dwarf galaxies. $(M/L_B)_{est}$ (last column of Table 2) represents the mass-to-light ratio calculated using the estimated total (baryonic + dark matter) mass for the group and the B-band luminosity for the group. In other words, the last column of Table 2 is simply the ninth column divided by the third column of that same table.

**Methods: Calculation of isolation fraction in SDSS** To estimate that < 5% of dwarf galaxies are observed to have close companions as stated in the first paragraph of the main text, we have used a stellar mass catalog[29] which is based on the SDSS Data Release 7 spectroscopic catalog[41] and a further value added catalog of improved photometry[42]. We then restricted this list to only low mass ($M_* < 5 \times 10^9$ $M_\odot$) galaxies that are at least ten times more massive than the limiting stellar mass of the sample at that redshift. In other words, we limited our analysis to those dwarf galaxies for which a potential lower mass neighbor with a stellar mass ratio as low as 1:10 would be detectable in the SDSS spectroscopic survey. The limiting stellar mass required for completeness at each redshift was determined using the conservative "red sequence + 3σ" fit[29], ensuring that the sample completeness is independent of galaxy color.

A total of 186 dwarf galaxies satisfy these criteria. As expected, these galaxies all lie within the low redshift portion of SDSS ($z < 0.015$). Due to the small sample size, we do not further restrict this sample based on proximity to the nearest massive galaxy. Of those 186 galaxies, only seven have companions both within a 2-D projected separation of 50 kpc and within a relative velocity difference of 225 km/s (parameters that define the maximum separations observed for the TNT groups). Each of the seven systems were visually inspected and confirmed to be genuine galaxy pairs. Thus only 3.8% of dwarf galaxies are observed to have close companions within the detection completeness limits of SDSS.

The observed fraction of galaxies with a close companion must be corrected to account for the overall spectroscopic incompleteness of the SDSS main galaxy sample, which is estimated to be ~12%[43]. This incompleteness leads to an underestimate of the fraction of galaxies which have close companions, but can be corrected for by applying statistical weights corresponding to the reciprocal of the spectroscopic completeness (1/0.88 in this case) to each galaxy[44]. This correction increases our estimate of the fraction of dwarfs with close companions from 3.8% to 4.3%, thus still below 5%.

We must also account for the additional small scale spectroscopic incompleteness that is present in SDSS due to fiber collisions. In particular, the inability to place two spectroscopic fibers closer together than 55″ biases SDSS against close spectroscopic galaxy pairs[45]. In principle, this source of incompleteness can be corrected for by applying statistical weights[46]. However, this source of incompleteness is negligible due to the very low redshifts of the dwarf galaxies under consideration here. In particular, a minimum angular separation of 55″ corresponds to a minimum projected physical separation of 6-16 kpc for the galaxies in our sample. Moreover, none of the detected close companions lie within these minimum separations. As such, no statistical correction to the observed fraction of dwarfs with close companions is warranted.

We address one final source of incompleteness by considering whether any of the 186 dwarfs lie within 50 kpc of the survey boundaries, thus leading to missed companions that fall outside of the SDSS spectroscopic coverage. However, estimates of the projected distance to the survey boundary for each galaxy in our SDSS catalog[47] reveal that none of the 186 dwarf galaxies in our sample are found to be within 50 kpc of the survey boundaries. As such, the survey boundaries have no effect on our measurement of the fraction of dwarf galaxies with close companions.

In conclusion, after taking into account all of the standard sources of incompleteness, we estimate that 95.7% of dwarf galaxies in SDSS have no companions within 50 kpc and 225 km/s and above a stellar mass ratio of 1:10. Assuming an uncertainty derived from Poisson statistics ($\sqrt{N}$), the resulting fraction of dwarf galaxies with close companions is observed to be 4.3 ± 1.6%.

We further note that the sample of 186 galaxies used in this calculation represents only 0.33% of the dwarf galaxies in our SDSS galaxy catalog. Thus, only a small fraction of dwarfs lie in the region of stellar mass-redshift parameter space where they and all of their close companions would be detectable by SDSS. Not only are dwarf pairs, and by extension dwarf groups, intrinsically rare, but they are also a challenge to detect given the completeness limits of current surveys. The combination of these two factors have thus required a sample as large as SDSS to reveal the first (albeit small) sample of dwarf galaxy groups presented here.

**Data Availability** The data that support the plots within this paper and other findings of this study are available from the corresponding author upon reasonable request.

**Acknowledgements** S. Stierwalt, S. E. Liss, and G. C. Privon thank S. Veilleux and M. McDonald for the use of their PI instrument, MMTF, and M. McDonald for sharing his advice and wisdom throughout the MMTF observations and data reduction. S. Stierwalt acknowledges the L'Oreal USA For Women in Science program for their grant to conduct this resesarch. S. E. Liss acknowledges support from a National Science Foundation Graduate Research Fellowship under Grant No. DDGE-1315231. S.E. Liss was also partially funded by a Virginia Space Grant Consortium Graduate STEM Research Fellowship and a Clare Boothe Luce Graduate Fellowship. D. R. Patton acknowledges a Discovery Grant from the Natural Sciences and Engineering Research Council (NSERC) of Canada which helped to fund this research. G. C. Privon was supported by a FONDECYT Postdoctoral Fellowship (No. 3150361). N. Kallivayalil is supported by the NSF CAREER award 1455260.

These results are based on observations obtained with the Apache Point Observatory 3.5-meter telescope, which is owned and operated by the Astrophysical Research Consortium. This work has also used catalogs and imaging from the SDSS. Funding for the SDSS and SDSS-II has been provided by the Alfred P. Sloan Foundation, the Participating Institutions, the National Science Foundation, the U.S. Department of Energy, the National Aeronautics and Space Administration, the Japanese Monbukagakusho, the Max Planck Society, and the Higher Education Funding Council for England. The SDSS Web Site is http://www.sdss.org/. The SDSS is managed by the Astrophysical Research Consortium for the Participating Institutions. The Participating Institutions are the American Museum of Natural History, Astrophysical Institute Potsdam, University of Basel, University of Cambridge, Case Western Reserve University, University of Chicago, Drexel University, Fermilab, the Institute for Advanced Study, the Japan Participation Group, Johns Hopkins University, the Joint Institute for Nuclear Astrophysics, the Kavli Institute for Particle Astrophysics and Cosmology, the Korean Scientist Group, the Chinese Academy of Sciences (LAMOST), Los Alamos National Laboratory, the Max-Planck-Institute for Astronomy (MPIA), the Max-Planck-Institute for Astrophysics (MPA), New Mexico State University, Ohio State University, University of Pittsburgh, University of Portsmouth, Princeton University, the United States Naval Observatory, and the University of Washington.

Based on observations obtained at the Gemini Observatory (Program ID: GN-2016A-Q-16) and processed using the Gemini IRAF package, which is operated by the Association of Universities for Research in Astronomy, Inc., under a cooperative agreement with the NSF on behalf of the Gemini partnership: the National Science Foundation (United States), the National Research Council (Canada), CONICYT (Chile), Ministerio de Ciencia, Tecnología e Innovación Productiva (Argentina), and Ministério da Ciência, Tecnologia e Inovacão (Brazil).

**Competing Interests** The authors declare that they have no competing financial interests.

**Correspondence** Correspondence and requests for materials should be addressed to Sabrina Stierwalt (email: sabrinas@virginia.edu).

| Name | $v_{opt}$ [km/s] | g (unc) | g - r | log($M_*$) [$M_\odot$] | $R_{proj,prim}$ [kpc] | Velocity source |
|---|---|---|---|---|---|---|
| dm1049+09a | 10037($\pm$3) | 16.28 (0.06) | 0.29 | 9.32 (0.1) | 0.0 | SDSS |
| dm1049+09b | 10064($\pm$2) | 16.19 (0.06) | 0.14 | 9.11 (0.1) | 47.7 | SDSS |
| dm1049+09c | 10277($\pm$1) | 18.72 (0.19) | 0.20 | 8.19 (0.2) | 65.0 | APO |
| dm1049+09d | 10141($\pm$1) | 19.44 (0.27) | 0.19 | 7.66 (0.3) | 67.0 | APO |
| dm1049+09e | 9911($\pm$1) | 20.39 (0.41) | -0.25 | 7.10 (0.3) | 24.9 | APO |
| dm1623+15a | 10280($\pm$2) | 17.47 (0.11) | 0.28 | 8.77 (0.1) | 0.0 | SDSS |
| dm1623+15b | 10031($\pm$2) | 17.59 (0.11) | 0.18 | 8.61 (0.1) | 42.7 | SDSS |
| dm1623+15c | 10004($\pm$1) | 16.71 (0.08) | 0.14 | 8.98 (0.1) | 58.1 | APO |
| dm1623+15d | 10139($\pm$1) | 17.62 (0.12) | 0.16 | 8.76 (0.1) | 78.6 | APO |
| dm1403+41a | 10580($\pm$2) | 16.73 (0.08) | 0.18 | 8.81 (0.1) | 0.0 | SDSS |
| dm1403+41b | 10619($\pm$2) | 17.52 (0.11) | 0.21 | 8.61 (0.1) | 22.1 | SDSS |
| dm1403+41c | 10319($\pm$2) | 16.63 (0.07) | 0.29 | 9.19 (0.1) | 22.9 | SDSS |
| dm1403+41d | 10370($\pm$1) | 17.11 (0.09) | 0.23 | 8.80 (0.1) | 16.7 | APO |
| dm1440+14a | 5429($\pm$3) | 16.54 (0.07) | 0.39 | 8.78 (0.1) | 0.0 | SDSS |
| dm1440+14b | 5423($\pm$2) | 17.60 (0.11) | 0.28 | 8.14 (0.1) | 46.5 | SDSS |
| dm1440+14c | 5381($\pm$1) | 18.30 (0.16) | 0.24 | 7.88 (0.2) | 32.0 | SDSS |
| dm1718+30a | 4446($\pm$2) | 15.56 (0.05) | 0.43 | 8.96 (0.1) | 0.0 | SDSS |
| dm1718+30b | 4428($\pm$2) | 17.00 (0.09) | 0.45 | 8.52 (0.1) | 30.9 | SDSS |
| dm1718+30c | 4569($\pm$1) | 17.54 (0.11) | 0.19 | 7.77 (0.1) | 4.6 | APO |
| dm0909+06a | 14021($\pm$2) | 17.22 (0.10) | 0.35 | 9.31 (0.1) | 0.0 | SDSS |
| dm0909+06b | 13799($\pm$2) | 17.53 (0.11) | 0.41 | 9.35 (0.1) | 32.2 | SDSS |
| dm0909+06c | 13704($\pm$1) | 19.52 (0.28) | 0.15 | 8.17 (0.3) | 31.8 | APO |
| dm1349-02a | 6913($\pm$3) | 17.33 (0.10) | 0.25 | 8.41 (0.1) | 0.0 | SDSS |
| dm1349-02b | 6988($\pm$2) | 18.15 (0.15) | 0.20 | 8.13 (0.2) | 14.3 | SDSS |
| dm1349-02c | 6898($\pm$1) | 18.91 (0.21) | 0.14 | 7.55 (0.3) | 15.8 | APO |

Table 1. **TNT Isolated Dwarf Group Candidate Members.** $v_{opt}$ is the line-of-sight velocity measured from the optical spectra derived from the source in the last column. $R_{proj,prim}$ is the 2D projected distance between each group member and the primary or most massive group member in kpc. The last column notes the source of the velocity given in the second column, either from the Sloan Digital Sky Survey spectroscopic catalog (SDSS) or from original observations presented here using the Dual Imaging Spectrograph at the Apache Point Observatory.

| Group Name | Dist (unc) [Mpc] | log($L_B$) (unc) [$L_\odot$] | log($M_*$) (unc) [$M_\odot$] | log($M_{HI}$) (unc) [$M_\odot$] | $\sigma_{3D}$ (unc) [km/s] | Size [kpc] | (M/$L_B$) bound [M/L]$_\odot$ | log(M) estimate [$M_\odot$] | (M/$L_B$) estimate [M/L]$_\odot$ |
|---|---|---|---|---|---|---|---|---|---|
| 1049+09 | 146 (±5) | 10.34 (±0.02) | 9.56 (±0.07) | 10.08 (±0.04) | 209 (±1) | 80.5 | >9.49 | 12.05 | 51.3 |
| 1623+15 | 147 (±5) | 10.19 (±0.03) | 9.40 (±0.05) | 10.29 (±0.03) | 188 (±2) | 78.4 | >10.46 | 11.92 | 53.8 |
| 1403+41 | 152 (±5) | 10.34 (±0.02) | 9.51 (±0.06) | 10.21 (±0.01) | 224 (±2) | 43.3 | >5.74 | 11.83 | 30.8 |
| 1440+14 | 78 (±4) | 9.51 (±0.03) | 8.91 (±0.08) | 9.56 (±0.03) | 37 (±2) | 32.2 | >0.79 | 10.60 | 12.3 |
| 1718+30 | 65 (±5) | 9.69 (±0.03) | 9.11 (±0.08) | 9.45 (±0.01) | 109 (±1) | 31.2 | >4.40 | 11.59 | 79.4 |
| 0909+06 | 201 (±5) | 10.13 (±0.04) | 9.65 (±0.07) | 9.55 (±0.06) | 230 (±2) | 32.3 | >7.45 | 11.83 | 50.1 |
| 1349-02 | 101 (±4) | 9.46 (±0.05) | 8.63 (±0.09) | 9.60 (±0.01) | 68 (±2) | 15.9 | >1.48 | 10.66 | 15.6 |

Table 2. **TNT Isolated Dwarf Group Candidates.** Adopted distances, total stellar masses, 3D velocity dispersions $\sigma_{3D}$, group sizes, estimated total masses $M_{est}$, and mass-to-light ratios required for the groups to be bound and estimated based on total mass content are calculated as described in Methods. Distance uncertainties are noted as the quadrature sum of the uncertainty due to the range of line-of-sight velocities observed for all group members and the uncertainty inherent in using Hubble flow to derive a distance assuming a peculiar velocity of 300 km/s. All other uncertainties noted in parentheses are calculated directly from the uncertainties in the measured quantities given in Table 1. Note that we include dm1349-02 in the TNT sample of isolated dwarf galaxy groups even though it is in fact marginally isolated. The nearest massive galaxy sits at 1.3 kpc away with a stellar mass of $10^{10.49}$ $M_\odot$.

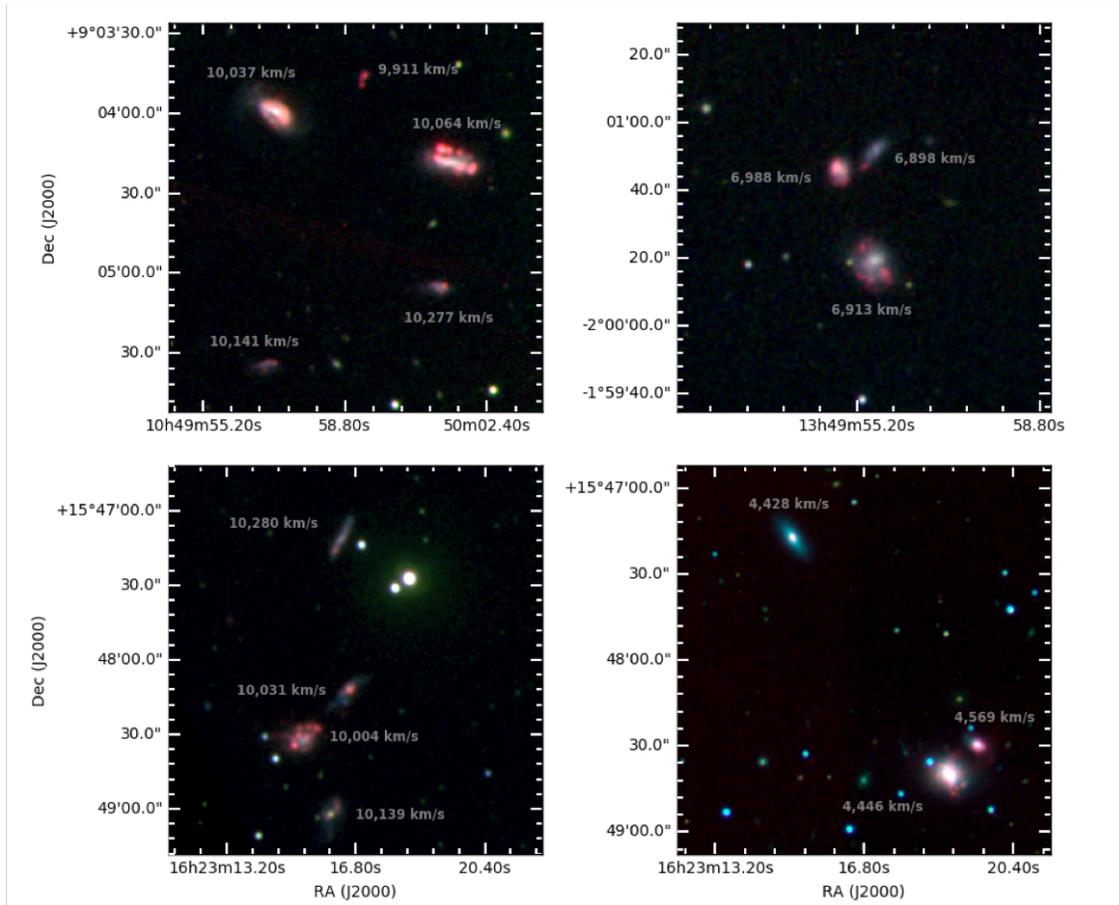

Figure 1. **TNT Dwarf Groups**. Three color images of four of the dwarf groups obtained with the Maryland Magellan Tunable Filter (MMTF) Fabry Perot (Δλ = 10 Å) for dm1049+09 (top left), dm1349-02 (top right), and dm1623+15 (bottom left) and with the Gemini Multi-Object Spectrograph (Δλ = 50 Å) for dm1719+30 (bottom right). Red corresponds to Hα, green corresponds to MMTF or Gemini R-band, and blue corresponds to SDSS g-band. Velocities derived from optical spectroscopy are noted for each group member.

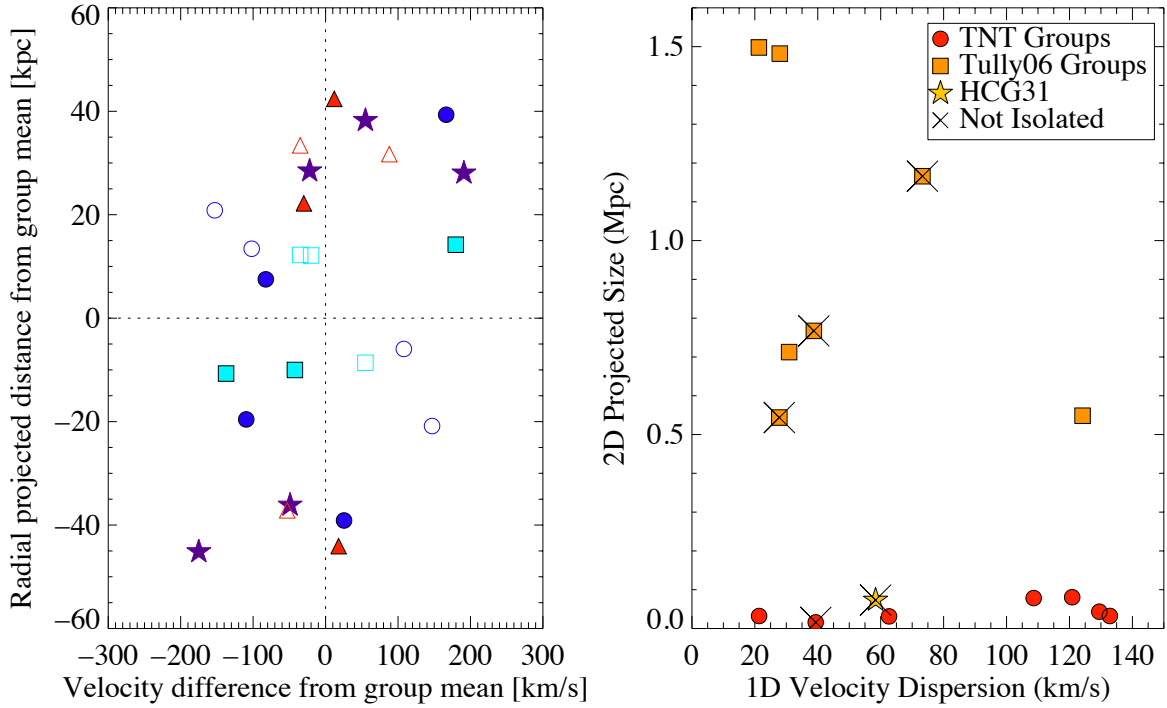

Figure 2. **Radial and Velocity Extent of TNT and Literature Dwarf Groups.** *Left:* Projected radial separation from the centroid of the group in kpc versus difference in group member line-of-sight velocity from the group mean in km/s. No weighting is applied. The groups are filled stars (dm1049+09), filled circles (dm1623+15); open circles (dm1403+41); filled triangles (dm1440+14); open triangles (dm1718+30); filled squares (dm0909+06); and open squares (dm1349-02). *Right:* 1D velocity dispersions and 2D projected sizes for the TNT dwarf groups presented here (red circles) and dwarf associations from the literature (orange squares and yellow star). Groups < 1.5 Mpc from a massive ($M_* \sim 10^{10}\,M_\odot$) host are marked with X's. In both panels, uncertainties are smaller than symbol sizes (see Table 1 for uncertainties in individual TNT velocity measurements).